\documentstyle[11pt,epsf]{article}

\def\indspace{\hspace*{1.0em} }
\def\appendix{\setcounter{section}{0}
\def\thesection{Appendix \Alph{section}}
\def\theequation{\Alph{section}.\arabic{equation}}}

\newfont{\subsub}{cmr6}

\newcounter{szk}

\begin{document}
\title{Pareto index induced from the scale of companies
}
\author{
\footnote{e-mail address: ishikawa@kanazawa-gu.ac.jp} Atushi Ishikawa
\\
Kanazawa Gakuin University, Kanazawa 920-1392, Japan
}
\date{}
\maketitle

\begin{abstract}
\indent
Employing profits data of Japanese companies in 2002 and 2003,
we confirm that Pareto's law and the Pareto index are derived from 
the law of detailed balance and Gibrat's law. 
The last two laws are observed beyond 
the region where Pareto's law holds.
By classifying companies into job categories,
we find that 
companies in a small scale job category
have more possibilities of growing
than those in a large scale job category.
This kinematically explains that the
Pareto index for the companies in the small scale job class is larger
than that for the companies in the large scale job class.
\end{abstract}
\begin{flushleft}
\end{flushleft}

\vspace{1cm}
\section{Introduction}
\label{sec-introduction}
\indspace
The pioneering discovery by Pareto \cite{Pareto} is not only an important cue for
research in fractal but also a significant issue in economics.
The Pareto law states that
a cumulative number distribution $N_P(> x)$ obeys power-law
for income $x$ which is larger than a certain observational threshold $x_0$:
\begin{eqnarray}
    N_P(> x) \propto x^{-\mu}~~~~{\rm for }~~~~x > x_0,
    \label{Pareto}
\end{eqnarray}
where the exponent $\mu$ is called Pareto index.
Recently Pareto's law is checked with high accuracy by using 
digitalized data \cite{ASNOTT}.
Although the number of persons or companies in
the range where Pareto's law (\ref{Pareto}) holds
is a few percent,
the amount of the income occupies a large part of the total income.
High income persons or high income companies can, therefore, influence economics.
It is quite important to investigate the mechanism which governs them.

The research on the income distribution is well investigated in econophysics \cite{MS}.
Many models to explain the distribution are proposed
(for instance recently Refs.~\cite{TTOMS, Mizuno, AIST, NS}).
Furthermore, Fujiwara et al. \cite{FSAKA} find that,
without assuming any model,
Pareto's law can be derived kinematically from
the law of detailed balance and Gibrat's law \cite{Gibrat}
which are observed in high quality digitalized data.
They derive not only Pareto's law but also the reflection law
which correlates a positive growth rate distribution to negative one
by the Pareto index $\mu$.
This means that Pareto's law and index are explained
by the law of detailed balance and Gibrat's law.
We should note that these arguments are limited in the high income region ($x > x_0$).

The findings in Ref.~\cite{FSAKA} are quite fascinating
especially in the idea that the
Pareto index can be understood by the growth rate distribution.
In this paper, we first investigate the application as follows.
The distribution of high income companies follows Pareto's law
and the Pareto index is about $1$.
This is called Zipf's law \cite{Zipf}.
It is also known that the distributions in most job categories follow Pareto's law.
Those Pareto indices, however, scatter around $1$ \cite{Mizuno, OTT, Ishikawa}.
We examine 
the reflection law that relates the change of the Pareto index 
to the change of the growth rate distribution in job categories.

There is, however, a problem in estimating the Pareto index by the growth rate distribution
in each job category.
The number of companies in several job classes
is not enough to observe the reflection law.
We clear this difficulty by employing profits data
which have no threshold in contrast to high income data
announced publicly.

By observing all positive profits data,
we confirm the law of detailed balance.
We also find that
Gibrat's law holds in the region 
where the only initial profits are conditioned to be larger than the threshold.
In this case, Pareto's law is also obtained
and the reflection law can be observed 
in companies data classified into job categories.
Using these arguments,
we confirm 
the correlation between the Pareto index 
and the growth rate distribution.

In addition,
we find that
companies in a small scale job category
have more possibilities of growing profits
than those in a large scale job category.
This kinematically explains that the
Pareto index for the companies in the small job class is larger
than that for the companies in the large scale job class,
which is reported in Ref.~\cite{Ishikawa}.

\section{Pareto's law derived from the detailed balance and Gibrat's law}
\label{sec-income}
\indspace
In this section, we briefly review the proof that the detailed balance and 
Gibrat's law lead to 
Pareto's law and the reflection law \cite{FSAKA}.

Let the income at two successive points in time
be denoted by $x_1$ and $x_2$.
Growth rate $R$ is defined as the ratio $R = x_2/x_1$.
The detailed balance and Gibrat's law can be summarized as follows.
\begin{itemize}
    \item  Detailed balance (Time-reversal symmetry)\\
            The joint probability distribution function (pdf)
            $P_{1 2}(x_1, x_2)$ is symmetric:
            \begin{eqnarray}
                P_{1 2}(x_1, x_2) = P_{1 2}(x_2, x_1).
            \label{Detailed balance}
            \end{eqnarray}
    \item  Gibrat's law\\
            The conditional probability distribution of growth rate $Q(R|x_1)$ 
            is independent of the initial value $x_1$: 
            \begin{eqnarray}
                Q(R|x_1) = Q(R),
            \label{Gibrat}
            \end{eqnarray}
            where $Q(R|x_1)$ is defined by using the pdf $P_1(x_1)$
            and the joint pdf $P_{1 R}(x_1, R)$ as
            \begin{eqnarray}
                Q(R|x_1) = \frac{P_{1 R}(x_1, R)}{P_1(x_1)}.
            \label{conditional}
            \end{eqnarray}
\end{itemize} 

\begin{figure}[htb]
 \centerline{\epsfxsize=0.8\textwidth\epsfbox{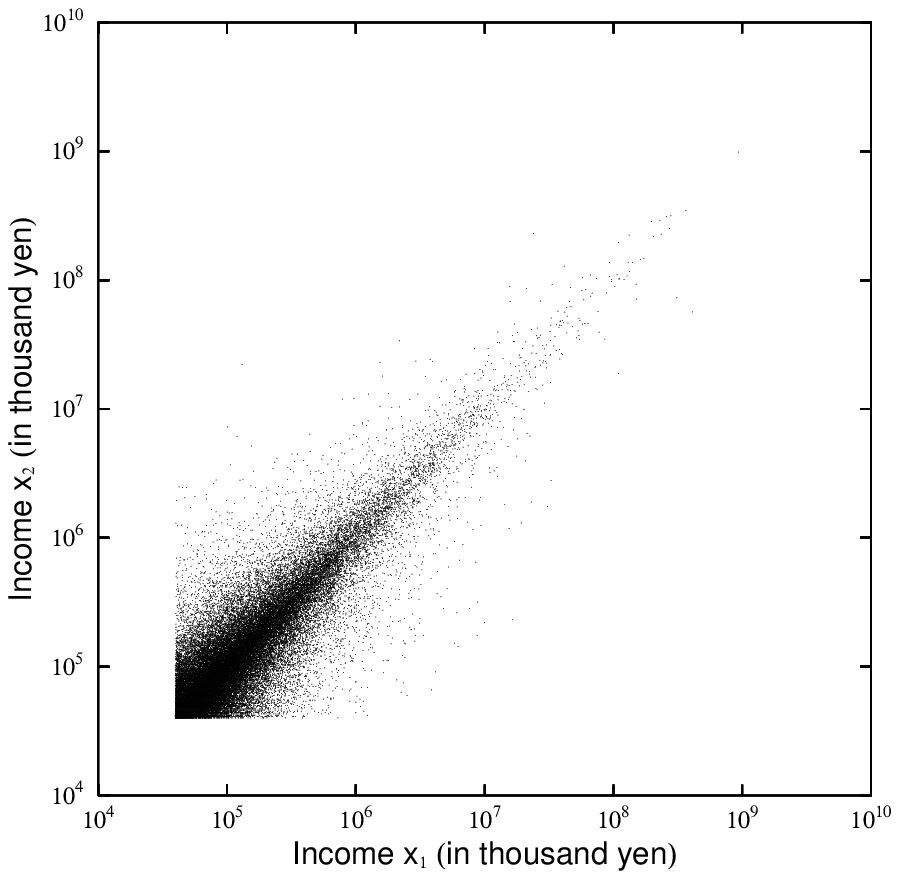}}
 \caption{The scatter plot of all companies, incomes of which in 2002 ($x_1$) and 2003 ($x_2$)
 exceeded 40 million yen ($x_0$), $x_1 > x_0$ and $x_2 > x_0$.}
 \label{Income2002vsIncome2003}
\end{figure}

\begin{figure}[htb]
 \centerline{\epsfxsize=0.8\textwidth\epsfbox{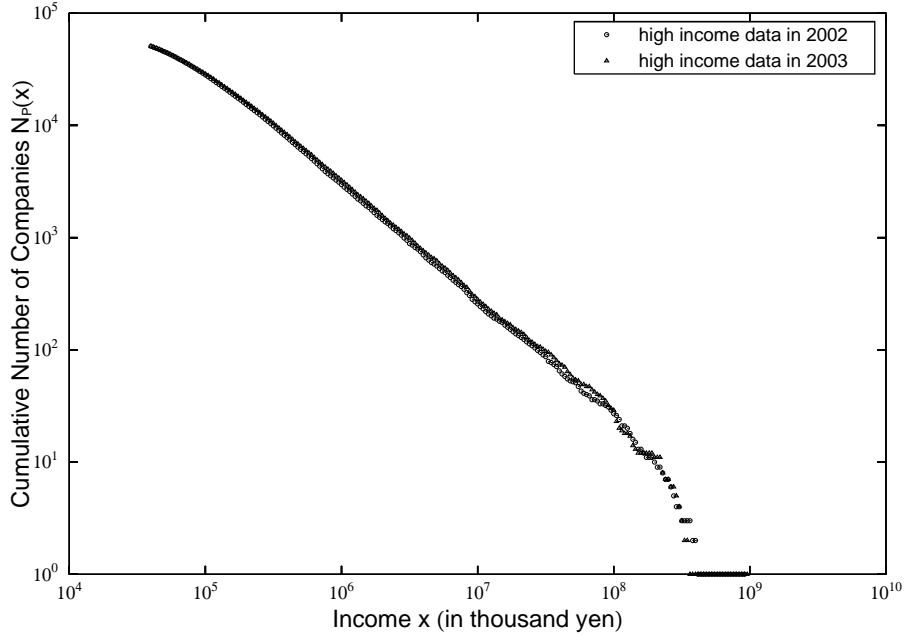}}
 \caption{Cumulative number distributions $N_P(x_1)$ and $N_P(x_2)$ for companies, the
 income of which in 2002 ($x_1$) and 2003 ($x_2$)
 exceeded $x_0$, $x_1 > x_0$ and $x_2 > x_0$ 
 ($x_0=4 \times 10^4$ thousand yen).}
 \label{IncomeDistribution}
\end{figure}

These phenomenological properties, for example, are observed 
in the data of high income Japanese companies in 2002 and 2003.
In Japan, companies having annual income more than 40 million yen
are announced publicly as ``high income companies'' every year.
The database is published by Diamond Inc.
In  Fig.~\ref{Income2002vsIncome2003},
all the companies are plotted, the
income of which in 2002 ($x_1$) and 2003 ($x_2$) exceeded 40 million yen ($x_0$):
\begin{eqnarray}
    x_1 > x_0 ~~~~~{\rm and} ~~~~~x_2 > x_0.
\label{income_condition}
\end{eqnarray}
The number of companies which satisfy this condition
is $50,632$.
In  Fig.~\ref{Income2002vsIncome2003},
the time-reversal symmetry (\ref{Detailed balance}) is apparent.
In Fig.~\ref{IncomeDistribution}, 
we confirm Pareto's law (\ref{Pareto}) for the companies, the
income of which satisfies the condition (\ref{income_condition}).
The Pareto indices $\mu$ are very close to $1$, and
the case is often referred to as Zipf's law \cite{Zipf}.

\begin{figure}[htb]
 \centerline{\epsfxsize=0.8\textwidth\epsfbox{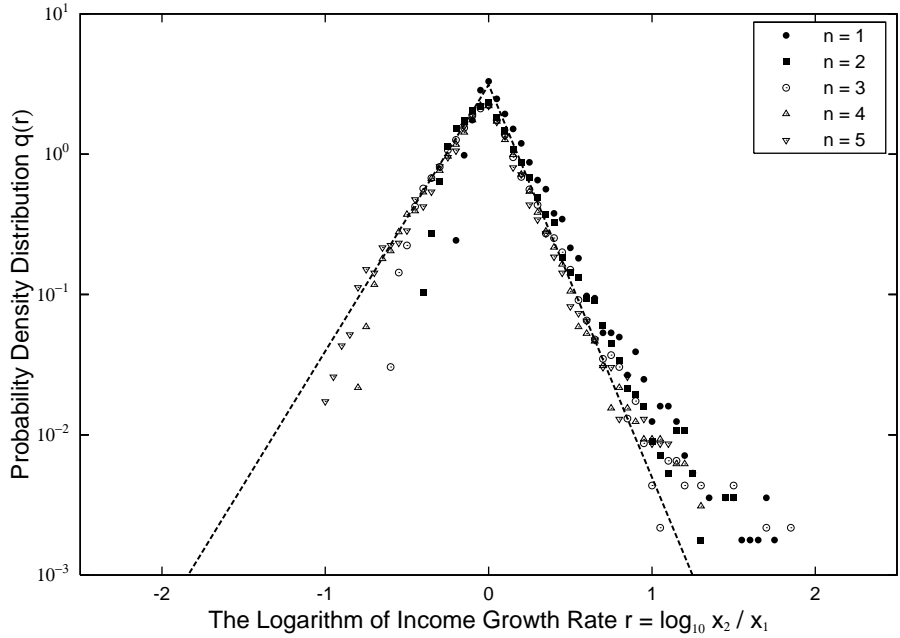}}
 \caption{The probability density distribution $q(r)$ of the log income growth rate
 $r \equiv \log_{10} x_2/x_1$ from 2002 to 2003.
 The data points are classified into five 
 bins of the initial income with equal magnitude in logarithmic scale,
 $x_1 \in 4 \times [10^{4+0.2(n-1)},10^{4+0.2n}]~(n=1,\cdots, 5)$.
 The data for large negative growth, $r \le 4+\log_{10} 4-\log_{10}x_1$,
 are not available because of the threshold, $x_2 > x_0$.}
 \label{IncomeGrowthRate}
\end{figure}

We also examine Gibrat's law in the regime (\ref{income_condition}).
Here we divide the range of $x_1$ into logarithmically equal bins as
$x_1 \in 4 \times [10^{4+0.2(n-1)},10^{4+0.2n}]$ with $n=1,\cdots, 5$.
In Fig.~\ref{IncomeGrowthRate},
the probability density for $r$ is expressed for each bin.
In all the cases, all the probability density distributions for different $n$
collapse on a single curve approximately.
This means that the distribution for the growth rate $r$
is approximately independent of the initial value $x_1$.
Here the probability density for $r$ defined by $q(r)$ is related to that for $R$ by
\begin{eqnarray}
    \log_{10}Q(R=10^r)+r+\log_{10}(\ln 10)=\log_{10}q(r).
\label{Qandq}
\end{eqnarray}

Pareto's law (\ref{Pareto}) can be derived
from the detailed balance (\ref{Detailed balance}) and 
Gibrat's law (\ref{Gibrat})
without assuming any model \cite{FSAKA}.
Due to the relation of
$P_{1 2}(x_1, x_2)dx_1 dx_2 = P_{1 R}(x_1, R)dx_1 dR$
under the change of variables from $(x_1, x_2)$ to $(x_1, R)$,
these two joint pdf's are related to each other,
\begin{eqnarray}
    P_{1 R}(x_1, R) = x_1 P_{1 2}(x_1, x_2).
\end{eqnarray}
By the use of this relation, the detailed balance (\ref{Detailed balance})
is rewritten in terms of $P_{1 R}(x_1, R)$ as follows:
\begin{eqnarray}
    P_{1 R}(x_1, R) = R^{-1} P_{1 R}(x_2, R^{-1}).
\end{eqnarray}
Substituting the joint pdf $P_{1 R}(x_1, R)$ for the conditional probability $Q(R|x_1)$
defined in Eq.~(\ref{conditional}),
the detailed balance is expressed as
\begin{eqnarray}
    \frac{P_1(x_1)}{P_1(x_2)} = \frac{1}{R} \frac{Q(R^{-1}|x_2)}{Q(R|x_1)}.
\end{eqnarray}

From the detailed balance and Gibrat's law (\ref{Gibrat}),
one finds the following:
\begin{eqnarray}
    \frac{P_1(x_1)}{P_1(x_2)} = \frac{1}{R} \frac{Q(R^{-1})}{Q(R)}.
\label{DandG}
\end{eqnarray}
By expanding this equation around $R=1$, the
following differential equation is obtained 
\begin{eqnarray}
    G'(1) P_1(x) + x P'_1(x) = 0,
\end{eqnarray}
where $x$ denotes $x_1$ and $G(R)$ is defined as $Q(R^{-1})/(R~Q(R))$.
The solution is given by
\begin{eqnarray}
    P_1(x) = C x^{-\mu-1}
\label{Pareto2}
\end{eqnarray}
with $G'(1) = \mu + 1$.
This is Pareto's law, which leads to the cumulative representation in Eq.~(\ref{Pareto})
as follows: $N_P(> x) = N_P(> x_0) P_1(> x) 
= N_P(> x_0) \int^{\infty}_{x} dy P_1(y) \propto x^{-\mu}$.

From Eqs.~(\ref{Pareto2}) and (\ref{DandG}),
one finds the relation between positive growth rate ($R>1$) and negative one ($R<1$),
\begin{eqnarray}
    Q(R) = R^{-\mu-2} Q(R^{-1}).
\label{reflection}
\end{eqnarray}
In terms of $q(r)$ in Eq.~(\ref{Qandq}),
this is written as
\begin{eqnarray}
    \log_{10}q(r)=-\mu r+\log_{10}q(-r).
\label{reflection2}
\end{eqnarray}
In Fig.~\ref{IncomeGrowthRate},
we examine this relation,
which is called the 'Reflection law' in Ref.~\cite{FSAKA}.
By the use of the reflection law to the best-fit line \cite{Mizuno}
\begin{eqnarray}
    \log_{10}q(r)=c-t_{+}~r~~~~~{\rm for}~~r > 0,
\label{line}
\end{eqnarray}
we obtain the line 
\begin{eqnarray}
    \log_{10}q(r)=c+t_{-}~r~~~~~{\rm for}~~r < 0
\label{rline}
\end{eqnarray}
with
\begin{eqnarray}
    \mu = t_{+} - t_{-},
\label{mu}
\end{eqnarray}
where $\mu \sim 1$.
The reflected line (\ref{rline}) seems to
match the data in Fig.~\ref{IncomeGrowthRate} \cite{FSAKA}.

\section{Pareto index induced
from the growth rate distribution}
\label{sec-profit}
\begin{figure}[htb]
 \centerline{\epsfxsize=0.8\textwidth\epsfbox{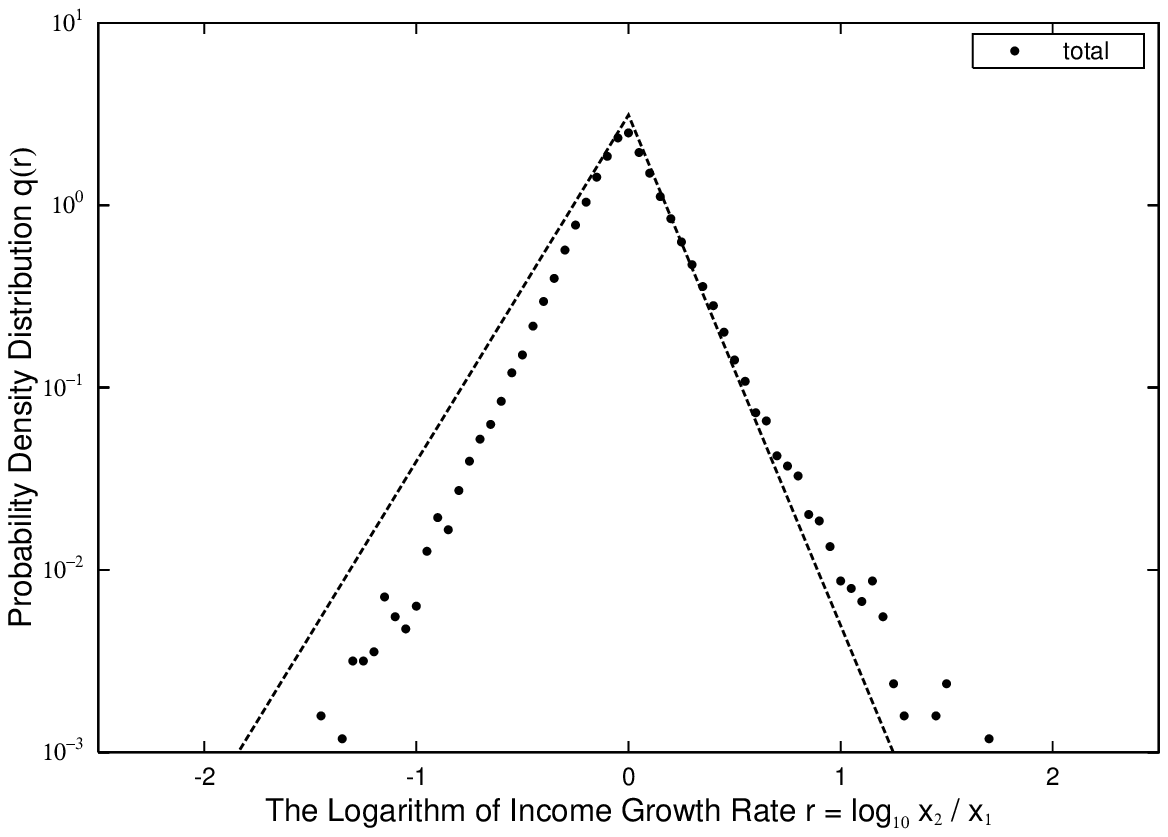}}
 \caption{The probability density distribution $q(r)$ of the log income growth rate $r$
 from 2002 to 2003 for all companies.}
 \label{IncomeGrowthRateTotal}
\end{figure}

\indspace
From the argument in the preceding section,
we expect the estimation of the Pareto index
by using Eq.~(\ref{mu})
under the detailed balance and Gibrat's law.
There is, however, a difficulty in measuring 
the slope $t_{-}$ of the reflected line (\ref{rline}).
The reason is that the data for large negative growth, 
$r \le 4 + \log_{10} 4 - \log_{10} x_1$,
are not available due to the limit (\ref{income_condition})
(Fig.~\ref{IncomeGrowthRate}).
The reflected line (\ref{rline}) does not match
the total data which are not classified into bins
for instance (Fig.~\ref{IncomeGrowthRateTotal}).
Indeed, the Pareto index is estimated to be about $0$ from
the best fit lines (\ref{line}) for $r>0$ and (\ref{rline}) for $r<0$.

\begin{figure}[htb]
 \centerline{\epsfxsize=0.8\textwidth\epsfbox{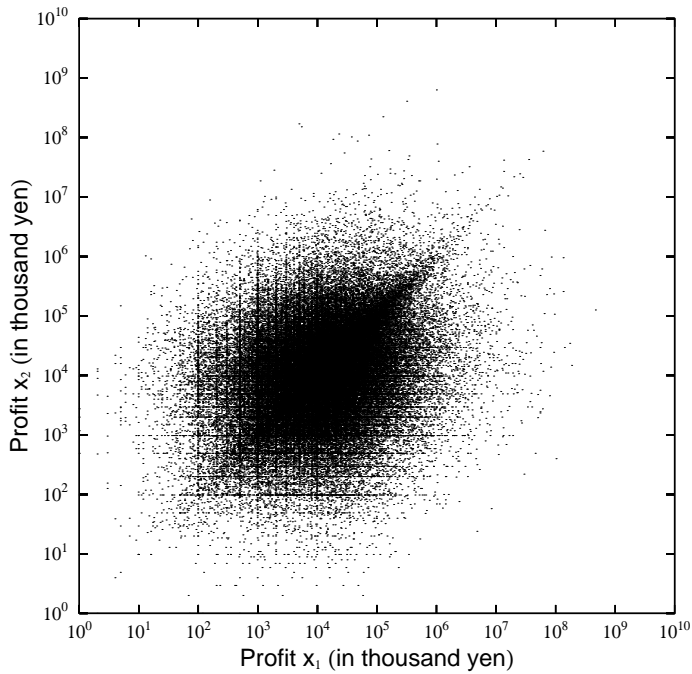}}
 \caption{The scatter plot of all companies, the profits of which in 2002 ($x_1$) and 2003 ($x_2$)
 exceeded $0$ yen, $x_1 > 0$ and $x_2 > 0$.}
 \label{Profit2002vsProfit2003}
\end{figure}

In order to avoid this problem,
we must remove the threshold $x_0$ in the region (\ref{income_condition}).
This is, however, impossible as long as we deal with the income data
announced publicly in Japan.
We employ profits as another economic quantity similar to the income.
Although profits of companies are not announced publicly,
the data are available on ``CD Eyes'' published by Tokyo Shoko Research, Ltd.
\cite{TSR}
for instance.
This database contains financial information of over 250 thousand companies in Japan.
We deal with the companies, the
profits of which in 2002 ($x_1$) and 2003 ($x_2$) exceeded 0 yen:
\begin{eqnarray}
    x_1 > 0 ~~~~~{\rm and} ~~~~~x_2 > 0.
\label{profit_condition}
\end{eqnarray}
In  Fig.~\ref{Profit2002vsProfit2003},
all the companies which satisfy this condition are plotted,
the number of which
is $132,499$.
The time-reversal symmetry (\ref{Detailed balance}) is apparent even
in Fig.~\ref{Profit2002vsProfit2003}.

\begin{figure}[htb]
 \centerline{\epsfxsize=0.8\textwidth\epsfbox{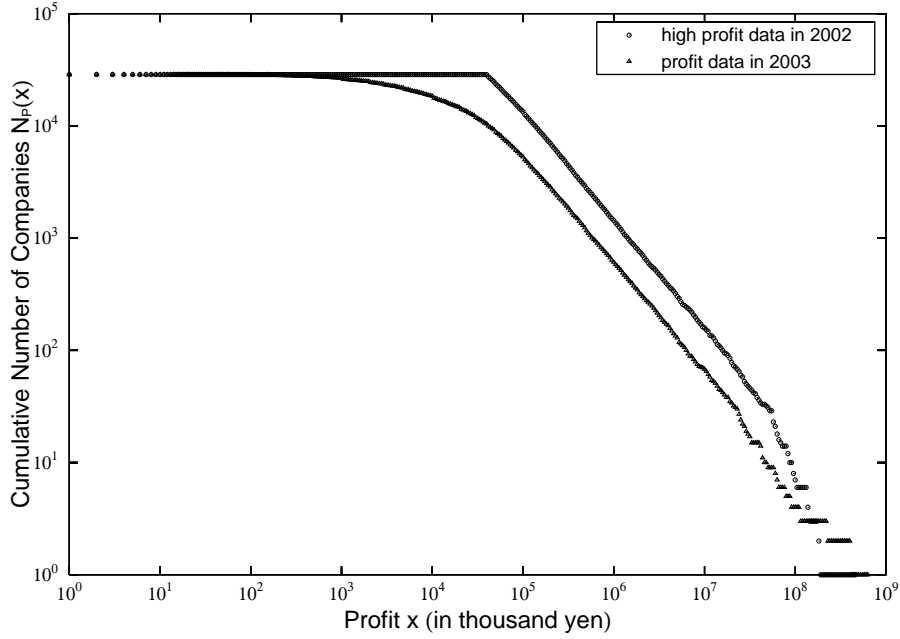}}
 \caption{Cumulative number distributions $N_P(x_1)$ and $N_P(x_2)$ for companies, the
 profits of which in 2002 ($x_1$) exceeded $x_0$
 and those in 2003 ($x_2$)
 exceeded $0$, $x_1 > x_0$ and $x_2 > 0$ 
 ($x_0=10^4$ thousand yen).
 Note that there is no threshold with respect to $x_2$.}
 \label{ProfitDistribution}
\end{figure}

\begin{figure}[htb]
 \centerline{\epsfxsize=0.8\textwidth\epsfbox{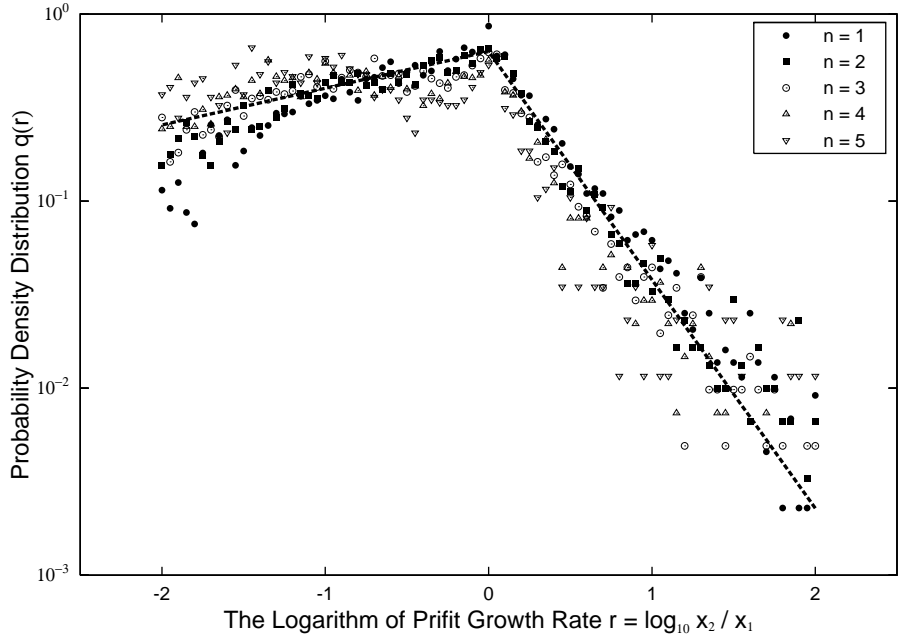}}
 \caption{The probability density distribution $q(r)$ 
 of the log profits growth rate
 $r \equiv \log_{10} x_2/x_1$ from 2002 to 2003.
 The data points are also classified into five 
 bins of the initial profits with equal magnitude in logarithmic scale:
 $x_1 \in 4 \times [10^{4+0.2(n-1)},10^{4+0.2n}]~(n=1,\cdots, 5)$.
 The data for large negative growth $r$
 are available in this case.}
 \label{ProfitGrowthRate}
\end{figure}

We desire to examine Gibrat's law without the threshold.
The condition is, therefore, employed:
\begin{eqnarray}
    x_1 > x_0 ~~~~~{\rm and} ~~~~~x_2 > 0,
\label{profit_condition2}
\end{eqnarray}
where $x_0$ is 40 million yen.
The number of the companies which satisfy this limit
is $28,644$.
In Fig.~\ref{ProfitDistribution},
Pareto's law (\ref{Pareto}) for the companies is observed
in the region (\ref{income_condition}), not in (\ref{profit_condition2}).
Let us examine Gibrat's law in the regime (\ref{profit_condition2}).
We classify the data for the high profits companies, $x_1 > x_0$,
into five bins in the same manner
in the preceding section.
The result is shown in Fig.~\ref{ProfitGrowthRate}.
We confirm Gibrat's law without the forbid region caused by the threshold.

Under the extended limit (\ref{profit_condition2}),
we can derive Pareto's law.
In the derivation, 
we should pay attention to the region where Gibrat's law holds as follows:
\begin{eqnarray}
    Q(R|x_1) &=& Q(R)~~~~~~~{\rm for}~~~~~x_1 > x_0~~{\rm and}~~x_2 > 0,\\
    Q(R^{-1}|x_2) &=& Q(R^{-1})~~~~{\rm for}~~~~~x_1 > 0~~~{\rm and}~~x_2 > x_0.
\label{ProfitGibrat}
\end{eqnarray}
In the region (\ref{income_condition}) 
where these two conditions are satisfying,
Eq.~(\ref{DandG}) 
is relevant and Pareto's law (\ref{Pareto2}) holds as a result.
This is observed in Fig.~\ref{ProfitDistribution}.

\begin{figure}[htb]
 \centerline{\epsfxsize=0.8\textwidth\epsfbox{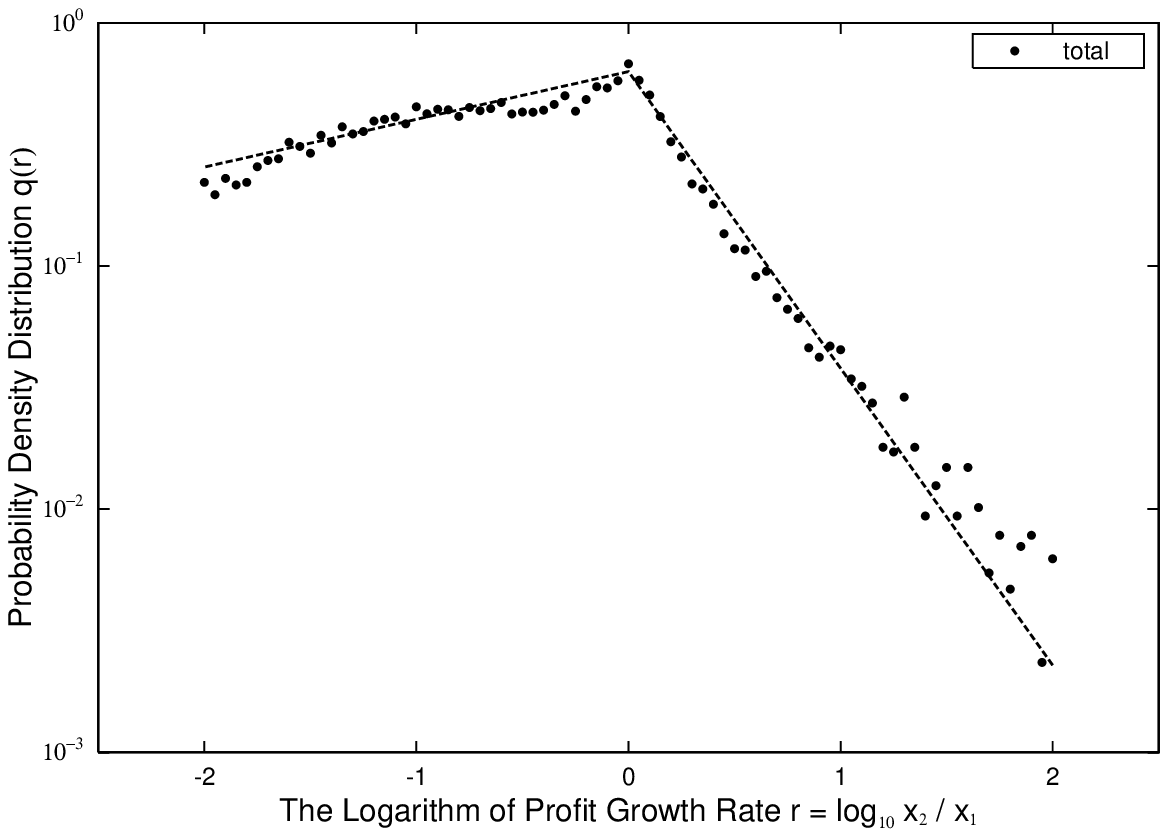}}
 \caption{The probability density distribution $q(r)$ of the log profit growth rate $r$
 from 2002 to 2003 for all companies.}
 \label{ProfitGrowthRateTotal}
\end{figure}

The reflection law (\ref{reflection}) is relevant
in the region (\ref{income_condition}) strictly.
Here we assume that Eqs.~(\ref{reflection}) and (\ref{reflection2})
hold approximately
near $R=1$ ($r=0$) even in the region (\ref{profit_condition2}),
the validity of which should be checked against the results.
Under this assumption, 
we evaluate the Pareto index by using Eq.~(\ref{mu}).
In Fig.~\ref{ProfitGrowthRateTotal}, 
the Pareto index is estimated to be $1.02 \pm 0.02$ from
the best fit lines (\ref{line}) for $r>0$ and (\ref{rline}) for $r<0$
for the total data which are not classified into bins.
This is consistent with the assumption.
We should comment that
the Pareto index is estimated to be about $0$ for the companies,
the profits of which in 2002 ($x_1$) and 2003 ($x_2$) exceeded $x_0$.

\section{High profits companies in job categories}
\label{sec-job}
\indspace
It is reported that income distributions in most job categories
follow Pareto's law, but those Pareto indices scatter around $1$
\cite{Mizuno, OTT, Ishikawa}. The
same phenomenon is observed in profits distributions.
In this section, we classify the companies into job categories
in order to verify the results in the preceding section.

In Japan, companies are categorized by 
Japan Standard Industrial Classification (JSIC).
This industrial classification is composed of four stages,
namely divisions, major groups, groups and details (industries).
The composition includes
14 divisions, 99 major groups, 463 groups and 1,322 industries.
The classification in the database ``CD Eyes'' follows JSIC.

We classify the $28,644$ companies into the 14 divisions, 
the profits of which satisfy the condition (\ref{profit_condition2}).
Because the number of companies classified into 99 major groups is not enough
to investigate the growth rate distribution.
The name of each division and the number of companies are follows,
A: Agriculture (84), B: Forestry (1), C: Fisheries (14), D: Mining (3,512), E: Construction (1,182),
F: Manufacturing (6,428), G: Electricity, Gas, Heat Supply and Water (218), 
H: Transport and Communications (2,300),
I: Wholesale, Retail Trade and Eating $\&$ Drinking Places (8,699),
J: Finance and Insurance (1,994), K: Real Estate (293), L: Services (3,919), M: Government, N.E.C. (0)
and N: Industries Unable to Classify (0).

\begin{figure}[htb]
 \centerline{\epsfxsize=0.8\textwidth\epsfbox{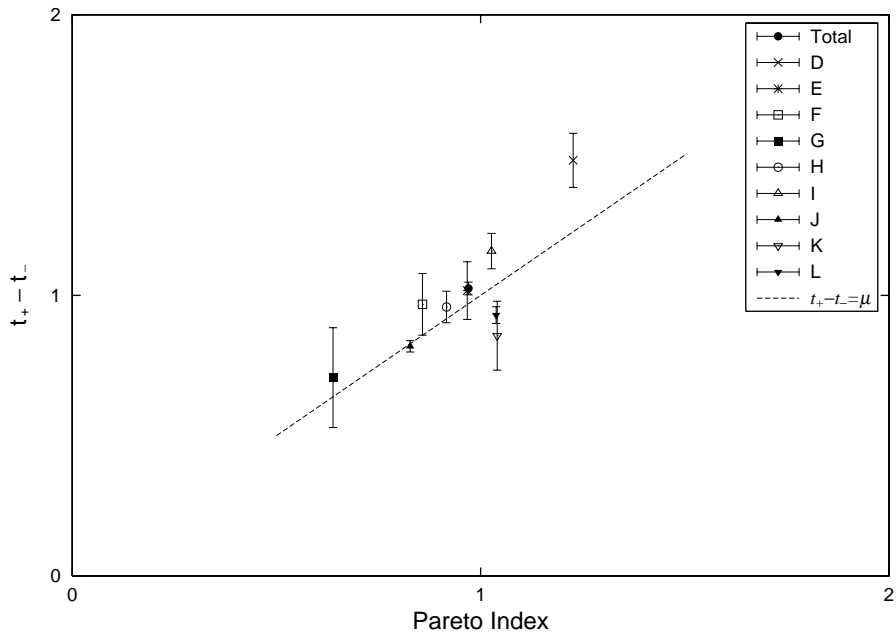}}
 \caption{Pareto indices directly measured in the power-law 
 and differences between $t_{+}$ and $t_{-}$  in job categories.
 In the legend, each alphabet represents the following division,
D: Mining, E: Construction,
F: Manufacturing, G: Electricity, Gas, Heat Supply and Water, H: Transport and Communications,
I: Wholesale, Retail Trade and Eating $\&$ Drinking Places,
J: Finance and Insurance, K: Real Estate and L: Services.}
 \label{ParetoIndexvsTau}
\end{figure}

In each job division, we measure $t_{+}$ and $t_{-}$ of the best fit lines 
(\ref{line}) and (\ref{rline})
of all the companies in the class. The
Pareto index can be estimated from the relation (\ref{mu})
by assuming the detailed balance and Gibrat's law.
Here we have excluded A, B, C, M and N divisions,
because the number of companies in each division is not enough to observe
the growth rate distribution.
By comparing the Pareto indices estimated by the growth rate with
those directly measured in the power-law,
we reconfirm the relation (\ref{mu}) 
in Fig.~\ref{ParetoIndexvsTau}.

After identifying the Pareto index change as 
the change of the growth rate distribution in job categories,
we consider an economic quantity which influences the growth rate.
In balance-sheet,
the sum of capital stock and retained earnings is equity capital,
and the sum of equity capital and debt is assets.
It is natural to consider that
a company works based on the assets and gets profits ($\sim$ income).
The positive correlation between assets and income is reported
in Ref.~\cite{OTT}.
It is also pointed out that assets and sales correlate positively
in the same reference.
On the other hand, the positive correlation between sales and capital stock
is observed in the database ``CD Eyes''.
Consequently there is a positive correlation between
assets and capital stock\footnote{
This correlation is probably caused by the Japanese commercial law.
In Japan, capital stock is treated as an important quantity
to protect creditors.}.
We employ, therefore, average ``capital stock'' as a quantity 
which statistically characterizes the scale of companies in each job category\footnote{
Indeed, the database ``CD Eyes'' includes no information about
equity capital or assets.}.

\begin{figure}[htb]
 \centerline{\epsfxsize=0.8\textwidth\epsfbox{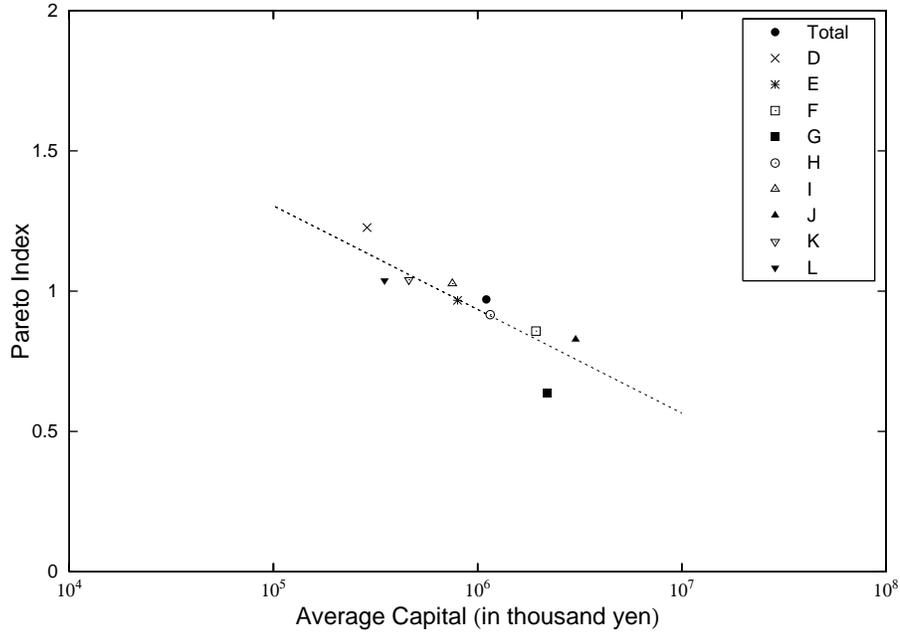}}
 \caption{The relation between the average capital and the Pareto index in job categories.
 In the legend, each alphabet represents the following division,
D: Mining, E: Construction,
F: Manufacturing, G: Electricity, Gas, Heat Supply and Water, H: Transport and Communications,
I: Wholesale, Retail Trade and Eating $\&$ Drinking Places,
J: Finance and Insurance, K: Real Estate and L: Services.}
 \label{AverageCapitalvsParetoIndex}
\end{figure}

In Ref.~\cite{Ishikawa},
it is reported that there is a relation
between the average capital stock of high income companies and the Pareto index
in 99 major groups.
The larger the average capital becomes, the smaller the Pareto index becomes.
The same phenomenon is observed in profits distributions (Fig.~\ref{AverageCapitalvsParetoIndex}).
This means that the Pareto index is related to the scale of companies in job categories.

\begin{figure}[htb]
 \centerline{\epsfxsize=0.8\textwidth\epsfbox{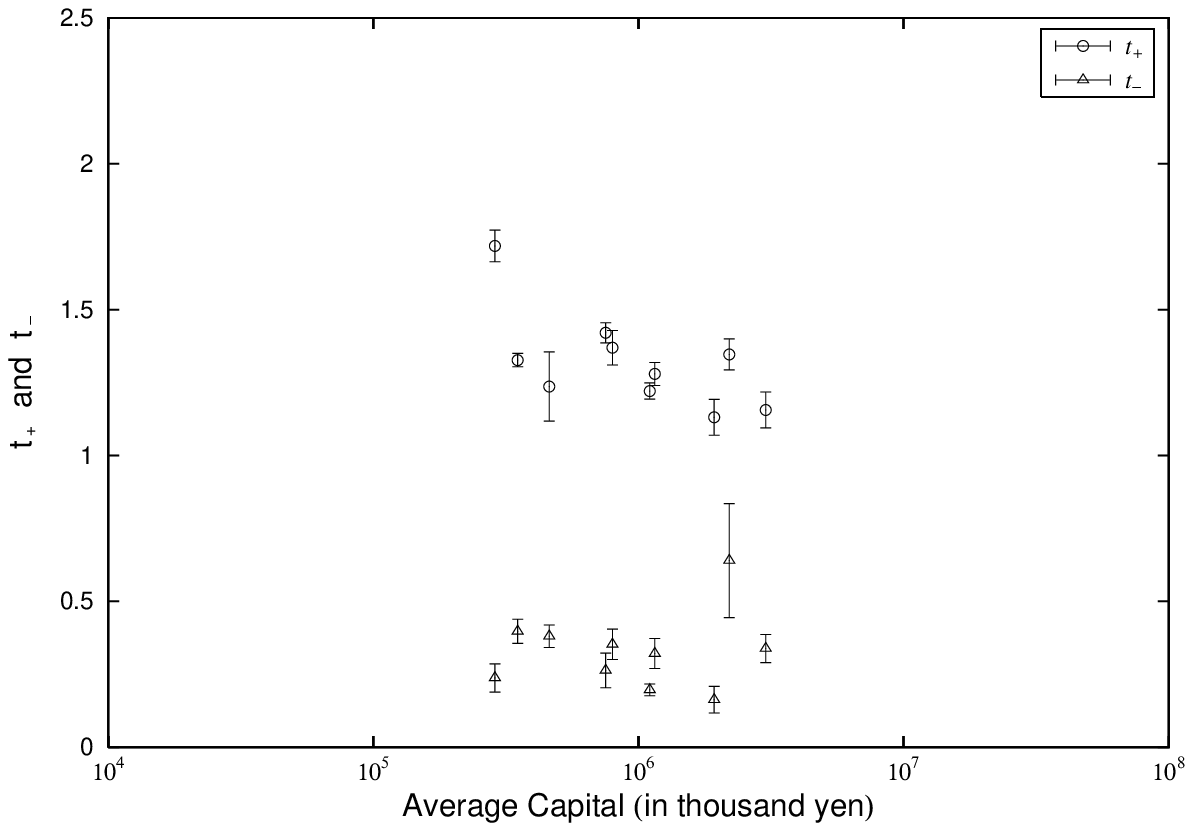}}
 \caption{The relation between the average capital and $t_{+}$, $t_{-}$ in job categories.
 One can identify the job category of each point by using the legend in Fig.\ref{AverageCapitalvsParetoIndex}.}
 \label{AverageCapitalvsTau}
\end{figure}

We can understand the kinematics
by the relation between the average capital and the slope of growth rate distribution
$(t_{+}, t_{-})$.
In Fig.~\ref{AverageCapitalvsTau},
$t_{+}$ decreases with the average capital
and $t_{-}$ hardly responds to that.
The companies in the small scale job category (Mining for instance)
have more possibilities of growing profits
than the companies in the large scale job category 
(Finance and Insurance for instance).
On the other hand,
the possibility of growing down 
is independent of the scale of the job category
for high profits companies, $x_1 \ge x_0$.
As a result, 
the relation (\ref{mu}) explains that the
Pareto index for the companies in the small scale job category 
is larger than that for the companies in the large scale job category.

\section{Conclusion}
\label{sec-Conclusion}
\indspace
In this paper,
we have first shown that the Pareto index can be estimated
by profits growth rate distribution of companies in Japan. 
The point is that the data have no threshold ($x_0$) in contrast to the high income data.
We find that the law of detailed balance and Gibrat's law hold in the extended region
$(x_1 > 0, ~x_2 > 0)$ and $(x_1 > x_0, ~x_2 > 0)$, respectively.

By classifying companies into 14 job divisions,
we have secondly confirmed the reflection law that relates the change of the Pareto index 
to the change of the growth rate distribution.

We thirdly find that the
companies in the small scale job category
have more possibilities of growing profits
than the companies in the large scale job category 
by observing the relation between the average capital and the slope of growth rate distribution.
The relation between the average capital and the Pareto index
is also verified.
Consequently it is kinematically explained that the
Pareto index for the companies in the small scale job category 
is larger than that for the companies in the large scale job category.

We have concentrated our attention to the region ($x_1 > x_0$, $x_2 > 0$)
in order to examine Gibrat's law.
Beyond this regime,
the breakdown of Gibrat's law is studied,
the recent works about that are done by Stanley's group 
\cite{Stanley1, Stanley2, Stanley3, Stanley4, Stanley5, Stanley6}.
In this paper, we find the law of detailed balance
over the regime where Pareto's law holds.
We expect that the distribution in the region ($x_1 \le x_0$, $x_2 > 0$)
can be derived kinematically
from the law of detailed balance and the results
found by Stanley's group.
This argument will kinematically decide the distribution for the middle region
where Pareto's law does not hold.


\section*{Acknowledgements}
\indent

The author is 
grateful to the Yukawa Institute for Theoretical 
Physics at Kyoto University,
where this work was initiated during the YITP-W-03-03 on
``Econophysics - Physics-based approach to Economic and
Social phenomena -'',
and specially to Professor~H. Aoyama for useful comments.
Thanks are also due to  Professor~H. Kasama
for careful reading of the manuscript.



\newpage



\begin{thebibliography}{99}
\bibitem{Pareto}
    V.~Pareto, Cours d'Economique Politique, Macmillan, London, 1897.
\bibitem{ASNOTT}
    H.~Aoyama, W.~Souma, Y.~Nagahara, H.P.~Okazaki, H.~Takayasu and M.~Takayasu, 
    cond-mat/0006038, Fractals 8 (2000) 293;\\
    W.~Souma, cond-mat/0011373, Fractals 9 (2001) 463. 
\bibitem{MS}
    R.N.~Mategna and H.E.~Stanley, An Introduction to Econophysics, 
    Cambridge University Press, UK, 2000.
\bibitem{TTOMS}
    H.~Takayasu, M.~Takayasu, M.P.~Okazaki, K.~Marumo and T.~Shimizu, cond-mat/0008057,
    in: M.M.~Novak (Ed.), Paradigms of Complexity, World Scientific, 2000, p. 243.
\bibitem{Mizuno}
    T.~Mizuno, M.~Katori, H.~Takayasu and M.~Takayasu, cond-mat/0308365,
    in: H.~Takayasu (Ed.), Empirical Science of Financial Fluctuations: The Advent of Econophysics, 
    vol. 321, Springer, Tokyo, 2003.
\bibitem{AIST}
    M.~Anazawa, A.~Ishikawa, T.~Suzuki and M.~Tomoyose,
    cond-mat/0307116, Physica A335 (2004) 616;\\
    A.~Ishikawa and T.~Suzuki, cond-mat/0403070, Physica A343 (2004) 376.
\bibitem{NS}
    M.~Nirei and W.~Souma, sfi/0410029.
\bibitem{FSAKA}
    Y.~Fujiwara, W.~Souma, H.~Aoyama, T.~Kaizoji and M.~Aoki,
    cond-mat/0208398, Physica A321 (2003) 598;\\
    H.~Aoyama, W.~Souma and Y.~Fujiwara, Physica A324 (2003) 352:\\
    Y.~Fujiwara, C.D.~Guilmi, H.~Aoyama, M.~Gallegati and W.~Souma,
    cond-mat/0310061, Physica A335 (2004) 197;\\
    Y.~Fujiwara, H.~Aoyama, C.D.~Guilmi, W.~Souma and M.~Gallegati,
    Physica A344 (2004) 112;\\
    H.~Aoyama, Y.~Fujiwara and W.~Souma, Physica A344 (2004) 117.
\bibitem{Gibrat}
    R.~Gibrat, Les inegalites economiques, Paris, Sirey, 1932.    \bibitem{Zipf}
    G.K.~Gipf, Human Behavior and the Principle of Least Effort, Addison-Wesley, Cambridge, 1949.
\bibitem{OTT}
    K.~Okuyama, M.~Takayasu and H.~Takayasu,
    Physica A269 (1999) 125.
\bibitem{Ishikawa}
    A.~Ishikawa, cond-mat/0409145, Physica A349 (2005) 597.
\bibitem{TSR}
    TOKYO SHOKO RESEARCH, LTD., http://www.tsr-net.co.jp/.
\bibitem{Stanley1}
    M.H.R.~Stanley, L.A.N.~Amaral, S.V.~Buldyrev, S.~Havlin, H.~Leschhorn, P.~Maass, M.A.~Salinger and H.E.~Stanley,
    Nature 379 (1996) 804.
\bibitem{Stanley2}
    L.A.N.~Amaral, S.V.~Buldyrev, S.~Havlin, H.~Leschhorn, P.~Maass,
    M.A.~Salinger, H.E.~Stanley and M.H.R.~Stanley, J. Phys. (France) I7 (1997) 621.
\bibitem{Stanley3}
    S.V.~Buldyrev, L.A.N.~Amaral, S.~Havlin, H.~Leschhorn, P.~Maass,
    M.A.~Salinger,  H.E.~Stanley and M.H.R.~Stanley, J. Phys. (France) I7 (1997) 635.
\bibitem{Stanley4}
    L.A.N.~Amaral, S.V.~Buldyrev, S.~Havlin, M.A.~Salinger and H.E.~Stanley, 
    Phys. Rev. Lett. 80 (1998) 1385.
\bibitem{Stanley5}
    Y.~Lee, L.A.N.~Amaral, D.~Canning, M.~Meyer and H.E.~Stanley,
    Phys. Rev. Lett. 81 (1998) 3275.
\bibitem{Stanley6}
    D.~Canning, L.A.N.~Amaral, Y.~Lee, M.~Meyer and H.E.~Stanley,
    Economics Lett. 60 (1998) 335.
\end{thebibliography}
\end{document}